# From Core to Periphery? Assessing Remote Work's Potential to Rebalance EU Regional Development

## Sławomir Kuźmar[1]

**Abstract**: The rapid expansion of remote work following the last pandemic has renewed interest in whether spatial decoupling of residence from workplace can contribute to rebalancing regional development across the European Union. This paper examines four interrelated dimensions of remote work-induced residential mobility using the R-MAP survey dataset, a large-scale cross-sectional survey of over 7,400 remote workers across Europe collected in 2024. First, the spatial direction of post-2020 relocations is analysed, revealing that mobility occurs overwhelmingly within the same urbanisation tier, with urban-to-urban moves accounting for 67% of all relocations. Counter-urban flows toward rural areas remain marginal at just 2% of moves, though their relative demographic impact on small rural populations is non-trivial. Second, the motivational structure of relocation decisions is examined, showing that quality-of-life considerations dominate (cited by 78% of movers), followed by economic and housing factors (70%), while digital infrastructure ranks among the least cited reasons. Third, amenity preferences are compared across residential contexts, documenting striking convergence between urban and rural remote workers, with statistically significant differences emerging only for public transport and restaurant access. Fourth, logistic regression models reveal that remote work intensity is a consistent positive predictor of relocation probability, with a transition from 50% to fully remote work associated with a 6.5 percentage point increase in relocation likelihood. Age, education, and industry sector also shape mobility patterns. Overall, the findings suggest that remote work primarily stretches metropolitan systems and reinforces peri-urban zones rather than triggering large-scale redistribution toward structurally weaker peripheral regions. Policy implications point toward targeted investment in rural amenities and service provision as a precondition for translating remote work potential into meaningful regional rebalancing.

[1]Assistant Professor, Department of Macroeconomics and Development Studies, Poznan University of Economics and Business, Poland, Al. Niepodległości 10, 61-875 Poznań, email: slawomir.kuzmar@ue.poznan.pl https://orcid.org/0000-0002-2458-0463

## Introduction

The COVID-19 pandemic has profoundly reshaped the spatial organisation of work across Europe. While remote work existed before 2020, the unprecedented scale of its adoption during lockdowns transformed it from a niche arrangement into a mainstream feature of contemporary labour markets (Aksoy et al., 2025). As hybrid and fully remote work arrangements have persisted well beyond the acute phase of the pandemic, a recurring question in both public discourse and policy debates is whether this shift can contribute to rebalancing regional development, potentially channelling population and economic activity from congested metropolitan cores toward structurally weaker peripheral regions.

This question arises against a backdrop of long-standing spatial inequalities within the European Union. Core metropolitan regions continue to concentrate high-productivity employment, skilled labour, and innovation capacity, while many peripheral and rural areas face demographic decline, brain drain, and underinvestment in services (Woźniak-Jęchorek & Kuźmar, 2025). Technological change has, in many respects, reinforced rather than mitigated these disparities, as Fiedler et al. (2021) show automation is positively associated with rising income inequality across Western European economies. Automation tends to concentrate gains in already-advantaged regions and skill groups. This raises a salient question: might remote work operate differently, enabling spatial redistribution of workers rather than further concentration of economic activity?

The pandemic has also altered the relationship between residence and workplace. Research documents significant, persistent declines in commuting. Surmařová et al. (2022) found a 38 percent decrease in passenger numbers on the Prague–Pilsen railway line between 2019 and 2020, with ridership not recovering even after restrictions were lifted and remote work becoming embedded in everyday routines. If commuting frequency declines durably, the economic logic tying residential location to workplace proximity weakens, potentially enabling residential

Yet whether such micro-level changes aggregate into meaningful regional rebalancing remains an open empirical question. The dominant narrative framing remote work as a catalyst for urban exodus and rural renaissance has outpaced the available evidence. Most post-pandemic research points to suburbanisation and metropolitan stretching rather than large-scale redistribution toward peripheral regions (Brueckner et al., 2023; Coskun et al., 2026). The extent to which remote work enables genuine spatial decoupling, as opposed to modest adjustments within existing metropolitan systems, depends on a complex interplay of individual preferences, housing market constraints, amenity provision, and institutional conditions.

This paper contributes to the emerging evidence base by examining four interrelated dimensions of remote work-induced residential mobility across Europe: the spatial direction of post-2020 relocations, the motivational structure underlying relocation decisions, the amenity preferences of urban versus rural remote workers, and the individual-level predictors of relocation probability. The analysis draws on the R-MAP survey dataset, a large-scale cross-sectional survey of over 7,400 remote workers across Europe collected in 2024 (Fellnhofer et al., 2025). By combining descriptive analysis of migration flows with logistic regression modelling, the study provides a multi-dimensional assessment of whether remote work is reshaping the European core–periphery landscape or primarily reinforcing existing spatial structures.

The remainder of the paper is organised as follows. Section 2 reviews the theoretical and empirical literature on remote work and residential mobility. Section 3 describes the data and methodology. Section 4 presents the results across all four analytical dimensions. Section 5 concludes with a discussion of policy implications and directions for future research.

## 1. Remote work and spatial mobility – literature review

Remote work's influence on spatial mobility has been studied since early telecommuting research (Nilles et al., 1976). Theoretical models based on the Monocentric City Model suggest that reducing commuting frequency lowers the marginal cost of distance, shifting residential equilibrium outward and facilitating longer-distance relocation (Lund & Mokhtarian, 1994; Shen, 2000; Rhee, 2009; Delventhal & Parkhomenko, 2025; Davis et al., 2021; Gokan et al., 2022).

However, empirical evidence remains mixed. Early studies found telecommuters concentrated in metropolitan areas rather than suburbs (Ellen & Hempstead, 2002), while others showed remote work often followed relocation rather than driving it (Ory & Mokhtarian, 2006). European studies found weak effects, with many using remote work to avoid moving (Muhammad et al., 2007; Ettema, 2010) or tolerating longer commutes (Zhu, 2013). Post-pandemic research on suburbanisation and the "donut effect" has yielded inconclusive findings (Ramani & Bloom, 2021; Ilham et al., 2024; Kuźmar et al., 2026). Some studies even suggest that remote work promotes residential immobility, enabling people to stay in place despite longer home-to-work distances (Hostettler Macias et al., 2025), or adopt complex arrangements, such as secondary homes near workplaces (Hostettler Macias et al., 2022).

Recent administrative and survey-based research has added further nuance to this picture. Evidence from Germany demonstrates that working from home increased work–home distances significantly after 2021, particularly for workers in occupations with high home-office potential, though most of this expansion occurred within the functional reach of large metropolitan labour markets rather than through long-distance migration to remote regions (Coskun et al., 2026). A similar pattern is observed in the metropolitan region of Madrid, where remote work increased moves from the core city toward suburban and peri-urban municipalities, particularly among young and middle-aged adults, without producing a large-scale exodus from the metropolitan system (Sánchez-Moral et al., 2026). These findings suggest that remote work stretches rather than dissolves metropolitan spatial structures.

At the same time, evidence from Belgium indicates that telework alone does not substantially increase the likelihood of residential relocation; life stage, household composition, and housing conditions remain dominant predictors of mobility decisions (Versighel et al., 2026). Complementary research from Sweden shows that more frequent working from home is associated with stronger preferences for neighbourhood amenities and proximity to everyday services rather than with detachment from urban environments altogether (Vilhelmson et al., 2026). Taken together, these findings point toward a process of spatial recalibration, in which remote work expands the radius of feasible residential locations while reinforcing demand for well-serviced, accessible environments, rather than a fundamental decoupling of residence from urban labour markets.

The ambiguity of these empirical findings is not merely a methodological artefact but reflects the inherent complexity of the relocation process itself. While remote work alters the economic logic of residential location, relocation remains a relatively rare and path-dependent event that cannot be reduced to a single trigger. Residential relocation theory emphasises that moves result not only from shifts in economic incentives but also from life-course transitions and household dynamics (Rossi, 1980; Mulder & Hooimeijer, 1999). Since Rossi's seminal work, triggers of residential mobility — marriage, childbirth, career changes — have been understood as events that generate a gap between current housing circumstances and desired living arrangements (Mulder, 1996). The life-course framework broadened this perspective by situating relocation decisions within parallel trajectories of family, education, labour, and housing, each capable of altering housing needs (Mulder & Hooimeijer, 1999; Coulter et al., 2016).

The stress–threshold model complements this perspective by explaining why changed preferences do not automatically translate into relocation (Brown & Moore, 1970). Households tolerate a degree of mismatch between their dwelling and evolving needs. Only when this mismatch exceeds a subjective threshold does relocation enter consideration, and even then the transition from intention to action depends on feasible alternatives, financial constraints, and market conditions.

Building on this theoretical foundation, Kuźmar (2026) proposed a schematic framework illustrating the residential mobility process in which remote work is integrated as a potential relocation trigger (Figure 1).

**Figure 1. Remote work as relocation trigger**

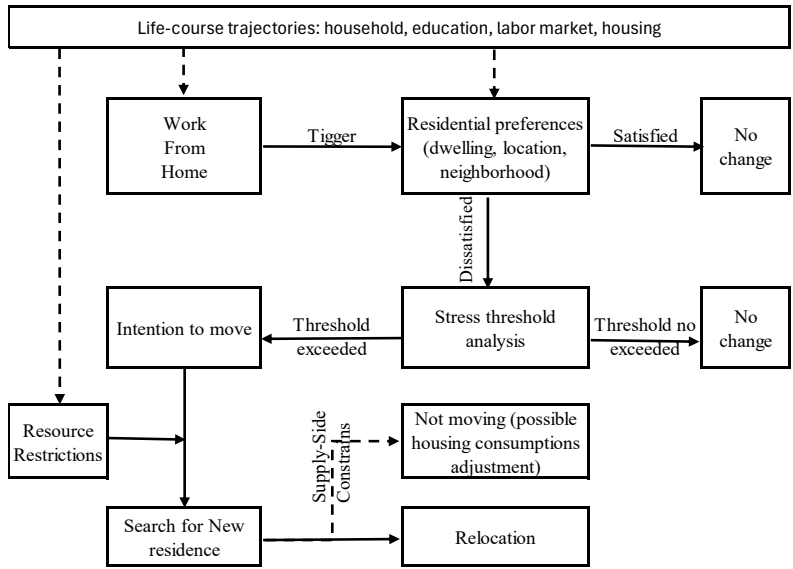

*Source: (Kuźmar, 2026).*

## 2. Data and Methodology

This study draws on the R-MAP survey dataset, a large-scale cross-sectional survey of remote workers across Europe. The survey was administered via the Prolific platform between July and August 2024 and collected responses from over 20,000 participants with European nationality. The questionnaire covered remote work perceptions and arrangements, residential location characteristics, amenity access, relocation practices, commuting patterns, and socio-demographic attributes. For details on the survey design and data collection methodology, see Fellnhofer et al. (2025).

From the full dataset, an analytical sample was constructed through multiple quality control filters (detailed procedures described in Appendix A1). These restrictions yield a working sample of 7,484 workers across Europe. The largest country samples come from the United Kingdom, Germany, Portugal, Poland, Italy, Spain, Netherlands, France, Hungary, and Greece. Table 1 presents the descriptive statistics for the analytical sample. The sample is relatively young, with a mean age of 34.3 years (SD = 9.9), and well-educated, reporting an average of 16.7 years of formal education (SD = 3.2). Respondents have an average of 12.4 years of work experience (SD = 9.8), though this variable shows the highest rate of missing data (9.4%). The sample is male-majority, with 55.4% male and 43.2% female respondents (with additional 1.1% non-binary and 0.2% preferring not to mention).

Work arrangements reflect the remote work focus of the study. On average, respondents work 22.5 hours per week remotely and 13.8 hours in the office, for a total of 36.2 hours per week. Work arrangement categories based on actual hours worked reveal that 23.7% work fully remotely (no office hours), 74.6% work in hybrid arrangements (combining remote and office hours), and only 1.7% work exclusively on-site. The three-category residential location typology (Urban, Intermediate, Rural), based on the degree of urbanisation classification, serves as the primary spatial dimension of the analysis. The sample is predominantly urban, comprising 5,626 urban residents (75.2%), 1,385 intermediate-area residents (18.5%), and 473 rural residents (6.3%). This distribution broadly reflects the urbanization patterns of contemporary Europe, though with some over-representation of urban areas typical of online panel recruitment.

Among the total sample, 1,206 respondents (16.1%) reported having relocated their residence since the onset of widespread remote work adoption in 2020. The share of relocated workers varies considerably across countries, reflecting both genuine differences in residential mobility patterns and potential country-specific selection effects in the Prolific platform recruitment process.

**Table 1. Descriptive Statistics**

*A. Continuous Variables*

| Variable | Mean | SD | Min | Max | N |
|---|---|---|---|---|---|
| Age (years) | 34.33 | 9.90 | 18 | 70 | 7,472 |
| Education (years) | 16.71 | 3.17 | 2 | 30 | 7,442 |
| Work experience (years) | 12.36 | 9.78 | 0 | 50 | 6,780 |
| Remote hours/week | 22.45 | 11.61 | 0 | 50 | 7,484 |
| Office hours/week | 13.75 | 11.43 | 0 | 50 | 7,484 |
| Total hours/week | 36.20 | 7.86 | 2 | 50 | 7,484 |

*B. Categorical Variables*

| Variable | Category | N | Percent |
|---|---|---|---|
| Gender | Male | 4,148 | 55.4 |
| | Female | 3,237 | 43.2 |
| Location type | Urban | 5,626 | 75.2 |
| | Intermediate | 1,385 | 18.5 |
| | Rural | 473 | 6.3 |
| Relocated since 2020 | Yes | 1,206 | 16.1 |
| | No | 6,278 | 83.9 |
| Work arrangement | Fully remote | 1,776 | 23.7 |
| | Hybrid | 5,581 | 74.6 |
| | Office only | 127 | 1.7 |

*Source: Authors' calculations based on R-MAP dataset (Fellnhofer et al., 2025).*

*Note: Missing values excluded from percentage calculations.*

This study examines four interrelated dimensions of remote work-induced residential mobility among European workers. The analysis proceeds in sequential steps, each addressing a distinct research question.

The first dimension concerns the spatial direction of relocation. While public discourse frequently frames remote work as a catalyst for urban exodus and counter-urbanisation mechanism, the empirical evidence remains mixed (Kuźmar et al., 2026). We examine whether relocated remote workers predominantly move toward less urbanised areas or whether relocation largely occurs within the same urbanisation tier. Migration directions are classified using a nine-category typology based on the degree of urbanisation classification (Urban, Intermediate, Rural), yielding directional flows from previous to current residential location. A simplified four-category migration pattern variable further distinguishes urban circulation from genuinely rural-oriented movements. The distribution of migration flows is examined descriptively, and the share of counter-urban movers (Urban→Rural) is assessed relative to the overall relocation population.

The second dimension addresses the motivational structure underlying relocation decisions. Respondents who relocated since 2020 were asked to indicate which of nineteen predefined reasons contributed to their move. We examine whether quality-of-life and economic considerations dominate over work-related factors, grouping individual reasons into four thematic categories: quality-of-life (encompassing life satisfaction, proximity to nature, safety, cultural immersion, and community), economic and housing (affordable housing, living expenses, housing size), work-related (workplace change, proximity to workplace, working conditions, digital infrastructure), and social and family factors (family and friends, children's schooling, health, retirement). The prevalence of each reason and thematic group is reported as the share of relocated workers citing the respective factor.

The third dimension examines whether residential location shapes amenity preferences among remote workers. Eight amenity categories are considered: groceries, healthcare, parks and green spaces, public transport, schools, restaurants and cafes, sports facilities, and cultural venues. For each amenity, a binary importance indicator is constructed, taking the value of one if the respondent considers the amenity important regardless of current access. Group differences between urban and rural residents are tested using two-sample t-tests, with statistical significance assessed at conventional thresholds.

The fourth and final dimension estimates the individual-level predictors of relocation using a series of binary logistic regression models. The dependent variable is an indicator of residential relocation since 2020. The key independent variable is remote work intensity, measured as the share of total weekly working hours spent working remotely. Three model specifications are estimated sequentially. Model 1 includes individual characteristics: age, gender, and years of formal education as well as previous residential location type. Model 2 extends this baseline with industry sector controls, with the ICT sector serving as the reference category given its dominant share in the sample and its status as the sector with the highest remote work penetration. Model 3 further incorporates country fixed effects to account for country-level heterogeneity in housing markets, labour mobility norms, and remote work adoption patterns. A unified previous location variable is constructed by combining observed pre-relocation location for relocated workers with current location for non-relocated workers, on the assumption that non-movers' previous and current locations coincide. Model results are reported as odds ratios, with average marginal effects computed for the preferred specification to facilitate direct interpretation of coefficient magnitudes in probability terms. Prior to model estimation, a near-perfect correlation between age and work experience ($r = 0.92$, $p < 0.001$) was identified, leading to the exclusion of work experience from all specifications to avoid multicollinearity. Discrimination ability is assessed using the area under the ROC curve (AUC) and model fit is compared across specifications using the Akaike and Bayesian Information Criteria.

## 3. Results

### 3.1 Spatial Directions of Relocation

Analysis of migration flows among the 1,206 relocated remote workers reveals a pattern that challenges the dominant narrative of remote work-driven urban exodus. As presented in Table 2 and illustrated, the overwhelming majority of relocations occur within the same urbanisation tier. Urban-to-urban moves dominate at 67.4% (n = 813). When all flows toward urban areas are combined, including moves from intermediate and rural origins, they account for 73.7% of all relocations (n = 889), underscoring that cities remain the primary destination of relocated remote workers rather than a source of out-migration. Counter-urban flows remain marginal in relative terms: urban-to-rural moves represent just 2.0% of all relocations (n = 24). However, when assessed relative to the size of destination populations, the picture is more nuanced. Urban-to-rural movers represent 5.1% of all current rural residents in the sample, a non-trivial inflow given that rural areas account for only 6.3% of the analytical sample. By contrast, rural-to-urban flows represent just 0.3% of urban residents, underscoring an asymmetry in the relative demographic impact of cross-tier mobility. The simplified migration pattern variable confirms the overall picture, with 91.0% of relocated workers classified as urban/intermediate circulation and only 2.7% (n = 33) as genuinely rural-oriented moves. These findings suggest that while remote work enables residential mobility primarily within existing urbanisation tiers, the relative impact of even small counter-urban flows on peripheral areas warrants careful consideration in regional policy discussions.

**Table 2. Directional flows of residential relocation by degree of urbanization**

| Migration Direction | Freq. | Relocation share (%) | Destination residents share (%) |
|---|---|---|---|
| Urban → Urban | 813 | 67.41 | 14.45 |
| Urban → Intermediate | 80 | 6.63 | 5.78 |
| Urban → Rural | 24 | 1.99 | 5.07 |
| Intermediate → Urban | 61 | 5.06 | 1.08 |
| Intermediate → Intermediate | 144 | 11.94 | 10.41 |
| Intermediate → Rural | 9 | 0.75 | 1.91 |
| Rural → Urban | 15 | 1.24 | 0.27 |
| Rural → Intermediate | 8 | 0.66 | 0.58 |
| Rural → Rural | 52 | 4.31 | 10.99 |
| Total | 1,206 | 100 | |

*Source: Author's calculations based on R-MAP dataset (Fellnhofer et al., 2025).*

*3.2 Motivational Structure of Relocation Decisions*

Figures 2 and 3 present the distribution of relocation reasons among previously urban (N = 917) and previously rural workers (N = 75), respectively. Across both groups, quality-of-life considerations emerge as the dominant motivational factor, cited by 67% of previously urban and 65% of previously rural relocated workers, the single most prevalent reason in both subsamples.

Among previously urban workers, quality-of-life is followed by a cluster of economic and housing factors: affordable housing (43%), housing size (39%), and living expenses (36%). Work-related reasons also feature prominently, with workplace change cited by 32% and working conditions by 24% of urban relocators. Family and friends (30%) and proximity to nature (22%) round out the top motivations, suggesting that remote work-enabled relocation among urban workers reflects a complex interplay of economic pressures, lifestyle aspirations, and social ties rather than a single dominant driver.

The motivational profile of previously rural workers shows notable similarities but also meaningful differences. Quality of life again leads (65%), but economic factors rank comparatively higher — living expenses (49%) and affordable housing (48%) occupy second and third positions respectively, suggesting that cost considerations may weigh more heavily among workers leaving rural areas. Family and friends (37%) and housing size (32%) follow, while proximity to nature (28%) — a reason more commonly associated with urban-to-rural moves — features prominently among rural movers as well, likely reflecting the heterogeneous nature of rural areas across Europe.

Work-related reasons, including workplace change, working conditions, and proximity to workplace, are each cited by approximately 21% of rural relocators, broadly consistent with the urban group. Digital infrastructure, by contrast, remains among the least cited reasons in both groups (7% urban, 4% rural), suggesting that connectivity concerns do not yet constitute a primary relocation barrier or motivator among remote workers in our sample.

Aggregating individual reasons into four thematic categories confirms the dominance of quality-of-life considerations (cited by 77.5% of all relocated workers), followed by economic and housing factors (70.1%), work-related reasons (54.3%), and social and family motivations (43.1%). These findings are consistent with the broader residential mobility literature, which emphasises the multi-determined nature of relocation decisions and the growing role of amenity and lifestyle factors in post-pandemic residential choices.

**Figure 2. Reasons for relocation among urban workers**

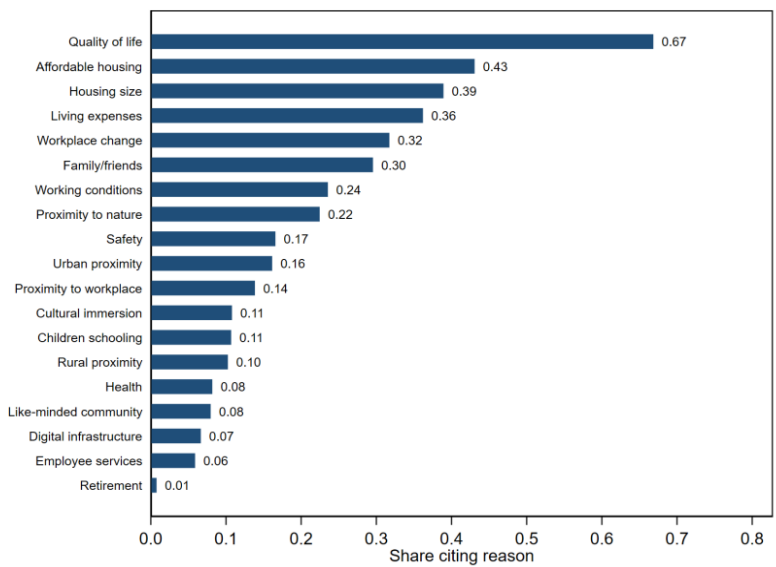

*Source: Authors' calculations based on R-MAP dataset (Fellnhofer et al., 2025).*

**Figure 3. Reasons for relocation among rural workers**

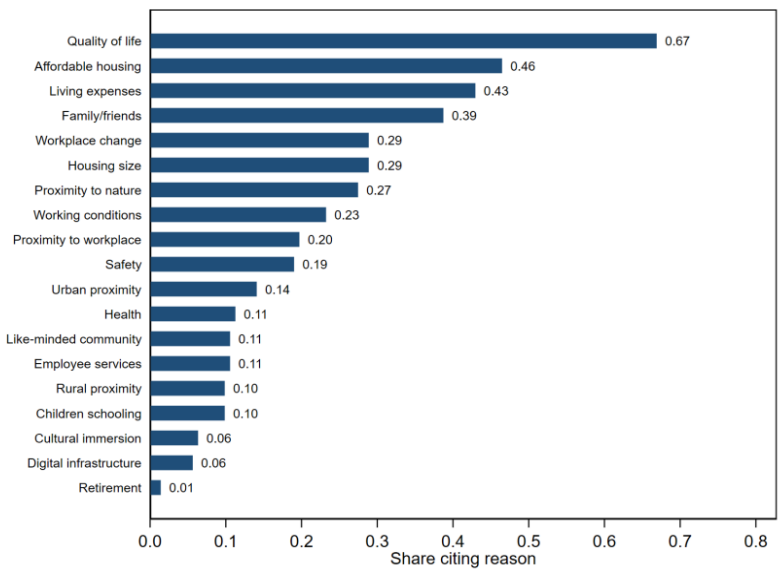

*Source: Authors' calculations based on R-MAP dataset (Fellnhofer et al., 2025).*

*3.3 Amenity Preferences: Urban versus Rural Remote Workers*

Figure 4 presents the share of urban and rural remote workers considering each of eight amenity categories important, regardless of current access. The comparison is restricted to urban (N = 5,626) and rural residents (N = 473), excluding intermediate-area workers from the group comparisons. The overall pattern reveals a striking degree of convergence across residential location types — urban and rural remote workers assign broadly similar importance to most amenities, with statistically significant differences emerging for only two categories.

As shown in Figure 4 and Table 3, the broader population of remote workers — the majority of whom have not yet relocated — represents a latent pool of potential movers whose amenity valuations may shape future residential mobility patterns. To the extent that rural residents already assign high importance to amenities typically associated with urban provision, such as healthcare and public transport, this signals unmet demand that could either motivate future outmigration from rural areas or, alternatively, be addressed through targeted rural investment to retain and attract remote worker populations.

The most consistently valued amenities across both groups are groceries (urban 69%, rural 67%), parks and green spaces (urban 68%, rural 67%), public transport (urban 71%, rural 60%), and healthcare (urban 59%, rural 63%). The high valuation of basic services such as groceries and healthcare across both groups suggests that remote workers, regardless of where they live, maintain strong preferences for access to essential infrastructure — a finding consistent with the quality-of-life orientation identified in the relocation reasons analysis.

Two statistically significant differences emerge from the t-test comparisons (Table 3). Public transport is valued significantly more by urban residents (71.3%) than by rural residents (60.0%), representing the largest observed gap across all amenities (t = 5.15, p < 0.001). This difference likely reflects both the greater availability of public transit in urban settings and the higher structural dependence of urban dwellers on collective mobility. Restaurants and cafes are also rated as more important by urban workers (39.0% vs. 33.0%, t = 2.58, p = 0.01), consistent with the role of urban social infrastructure in shaping everyday life preferences. A marginally significant difference is observed for healthcare, which rural residents value somewhat more highly (63.2% vs. 59.2%, p = 0.086), plausibly reflecting awareness of documented gaps in rural healthcare provision across Europe.

The absence of statistically significant differences for groceries, parks, schools, cultural venues, and sports facilities points to a broadly shared amenity preference structure among remote workers across the urban-rural continuum. This convergence is noteworthy in the context of regional development policy: if remote workers in rural areas value the same amenities as their urban counterparts, targeted investments in rural service provision may be sufficient to support and sustain remote worker populations outside major urban centres.

**Figure 4. Amenity Importance: Urban vs Rural Remote Workers**

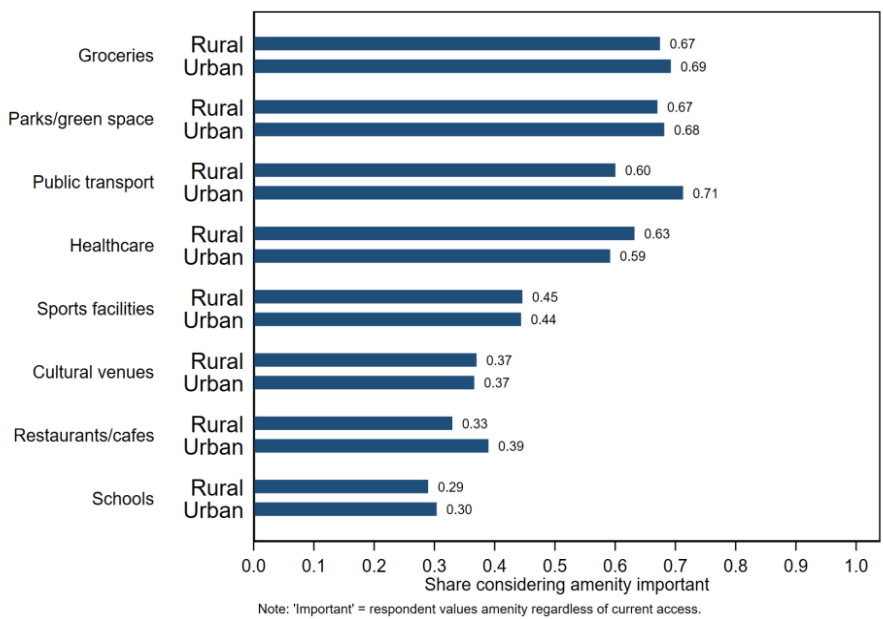

Note: 'Important' = respondent values amenity regardless of current access.

*Source: Authors' calculations based on R-MAP dataset (Fellnhofer et al., 2025)*
*Note: \*\*\* p<0.01, \*\* p<0.05, † p<0.10. Urban N = 5,626, Rural N = 473.*

**Table 3. Amenity Importance: Urban vs Rural Remote Workers (t-test results)**

| Amenity | Urban mean | Rural mean | Difference | t-stat | p-value |
|---|---|---|---|---|---|
| Public transport | 0.713 | 0.600 | 0.112 | 5.151 | 0.000*** |
| Restaurants/cafes | 0.390 | 0.330 | 0.060 | 2.576 | 0.010** |
| Healthcare | 0.592 | 0.632 | -0.040 | -1.720 | 0.086† |
| Groceries | 0.692 | 0.674 | 0.018 | 0.817 | 0.414 |
| Parks/green space | 0.681 | 0.670 | 0.011 | 0.506 | 0.613 |
| Sports facilities | 0.444 | 0.446 | -0.002 | -0.095 | 0.924 |
| Cultural venues | 0.366 | 0.370 | -0.004 | -0.166 | 0.868 |
| Schools | 0.304 | 0.290 | 0.014 | 0.642 | 0.521 |

*Source: Authors' calculations based on R-MAP dataset (Fellnhofer et al., 2025)*
*Note: \*\*\* p<0.01, \*\* p<0.05, † p<0.10. Urban N = 5,626, Rural N = 473.*

*3.4 Individual-Level Predictors of Residential Relocation*

Table 4 presents results from three logistic regression models estimating the probability of residential relocation since 2020. Model 1 includes individual characteristics and previous location type. Model 2 extends the specification with industry sector controls, addressing potential confounding through differences in remote work feasibility across sectors. Model 3 further adds country fixed effects to account for country-level heterogeneity in housing markets, labour mobility, and remote work culture. Discrimination ability, measured by the area under the ROC curve, improves progressively across specifications from AUC = 0.611 in Model 1 to AUC = 0.624 in Model 2 and AUC = 0.649 in Model 3 confirming that each additional set of controls contributes meaningfully to model performance. Model comparison based on the Akaike Information Criterion similarly favours Model 3 (AIC = 6,268 vs. 6,318 in Model 1), though the more conservative Bayesian Information Criterion penalises the larger parameter count and favours the more parsimonious Model 1 (BIC = 6,369). Given the theoretical motivation for country-level controls and the stability of key coefficients across specifications, Model 3 is retained as the preferred specification.

Remote work intensity emerges as a consistent and statistically significant predictor of relocation across all three specifications. Each one-percentage-point increase in remote work share raises the odds of relocation by a factor of 1.009 ($p < 0.001$) in all models — a remarkably stable estimate across alternative configurations. Expressed as an average marginal effect, this translates to a 0.13 percentage point increase in the probability of relocation per unit increase in remote work share (AME = 0.0012, $p < 0.001$). While the magnitude of this effect is modest at the individual level, it is substantively meaningful in aggregate: a worker transitioning from a 50% to a fully remote arrangement faces an estimated 6.5 percentage point higher probability of relocating relative to the baseline relocation rate of 16.1%.

Age is negatively associated with relocation across all models (OR = 0.959, $p < 0.001$), with each additional year reducing relocation probability by approximately 0.5 percentage points. Education shows a positive and stable effect (OR $\approx$ 1.027, $p < 0.05$), consistent with selective residential mobility. Gender is marginally significant (OR $\approx$ 1.14, $p < 0.10$), with female workers showing slightly higher relocation probability, though this finding is sensitive to specification. Previous location type shows no significant association, suggesting that relocation propensity is largely independent of prior residential context.

Among industries, workers in administrative support (OR = 0.554, $p < 0.001$), healthcare (OR = 0.634, $p < 0.05$), and other services (OR = 0.685, $p < 0.001$) show significantly lower relocation probabilities relative to the ICT reference category, while finance is marginally significant (OR = 0.778, $p < 0.10$). These differences likely reflect variation in remote work feasibility across sectors.

The overall explanatory power is modest (Pseudo $R^2$ = 0.044), consistent with residential mobility models where unobserved factors — housing markets, household composition, contractual arrangements — account for substantial variance.

## Table 4. Predictors of Residential Relocation

|  | (1) | (2) | (3) |
|---|---|---|---|
|  | Model 1 | Model 2 | Model 3 (Country FE) |
| Remote work share (%) | 1.009*** | 1.009*** | 1.009*** |
|  | (0.001) | (0.001) | (0.001) |
| Age (years) | 0.965*** | 0.964*** | 0.959*** |
|  | (0.003) | (0.004) | (0.004) |
| Female | 1.103 | 1.139* | 1.143* |
|  | (0.072) | (0.077) | (0.078) |
| Education (years) | 1.027** | 1.025** | 1.028** |
|  | (0.011) | (0.011) | (0.011) |
| Intermediate (prev. location) | 0.935 | 0.933 | 1.005 |
|  | (0.079) | (0.080) | (0.093) |
| Rural (prev. location) | 1.003 | 1.005 | 1.114 |
|  | (0.135) | (0.136) | (0.164) |
| Admin Support |  | 0.592*** | 0.554*** |
|  |  | (0.104) | (0.101) |
| Agriculture & Mining |  | 0.653 | 0.602 |
|  |  | (0.309) | (0.294) |
| Arts/Entertainment |  | 1.063 | 1.004 |
|  |  | (0.156) | (0.149) |
| Construction & Utilities |  | 0.977 | 0.981 |
|  |  | (0.171) | (0.174) |
| Education |  | 1.044 | 0.996 |
|  |  | (0.132) | (0.128) |
| Finance |  | 0.800* | 0.778* |
|  |  | (0.107) | (0.105) |
| Health |  | 0.672** | 0.634** |
|  |  | (0.117) | (0.112) |
| Manufacturing |  | 0.827 | 0.839 |
|  |  | (0.151) | (0.154) |
| Other Services | | 0.679*** | 0.685*** |
| | | (0.092) | (0.094) |
| Professional Services | | 0.889 | 0.889 |
| | | (0.098) | (0.099) |
| Public Sector | | 0.805 | 0.759 |
| | | (0.133) | (0.128) |
| Real Estate | | 0.768 | 0.736 |
| | | (0.307) | (0.299) |
| Retail & Hospitality | | 1.150 | 1.119 |
| | | (0.201) | (0.196) |
| Transportation | | 1.061 | 1.057 |
| | | (0.244) | (0.246) |
| Other/Missing | | 0.595** | 0.591** |
| | | (0.130) | (0.131) |
| Observations | 7331 | 7331 | 7327 |
| Log-likelihood | -3153.20 | -3136.82 | -3089.10 |
| AIC | 6320.40 | 6317.64 | 6268.19 |
| BIC | 6368.70 | 6469.44 | 6578.66 |
| Pseudo R² | 0.025 | 0.030 | 0.044 |

*Source: Authors' calculations based on R-MAP dataset (Fellnhofer et al., 2025)*

*Note: Odds ratios reported. Robust standard errors in parentheses. \*\*\* p<0.01, \*\* p<0.05, \* p<0.1.Reference categories: Urban location, ICT sector. Country fixed effects included in Model 3 but not reported.*

**Conclusion**

This paper examined four interrelated dimensions of remote work-induced residential mobility across Europe, drawing on a large-scale survey of over 7,400 remote workers collected in 2024. The findings suggest that remote work primarily stretches existing metropolitan spatial structures rather than triggering the large-scale core-to-periphery redistribution that public discourse frequently anticipates.

The analysis of migration flows reveals that post-2020 relocations occur overwhelmingly within the same urbanisation tier, with urban-to-urban moves accounting for 67% of all relocations and counter-urban flows from cities to rural areas representing just 2%. However, when assessed relative to destination population sizes, even these small flows constitute a non-trivial demographic inflow into rural communities, indicating that re-gional significance should not be evaluated solely in absolute terms. The motivational structure of relocation decisions confirms the multi-determined nature of residential mobility: quality-of-life considerations dominate, followed by economic and housing factors, while digital infrastructure ranks among the least cited reasons, suggesting that connectivity gaps do not yet represent the binding constraint on remote workers' location choices.

A particularly noteworthy finding concerns the striking convergence in amenity preferences between urban and rural remote workers, with significant differences emerging only for public transport and restaurant access. This implies that the challenge for peripheral regions is not one of attracting a fundamentally different type of resident but rather of matching the service expectations that remote workers hold regardless of current location. The logistic regression analysis further identifies remote work intensity as a consistent predictor of relocation: a transition from 50% to fully remote work is associated with a 6.5 percentage point increase in relocation probability, a substantively meaningful effect given the baseline relocation rate of 16.1%.

These results carry clear policy implications. Remote work alone is unlikely to serve as an autonomous mechanism of spatial convergence. Consistent with the stress–threshold model of residential mobility, remote work alters preferences and expands the radius of feasible residential locations, but actual relocation remains mediated by structural constraints — housing markets, hybrid office attendance requirements, and local amenity provision. Translating remote work potential into meaningful peripheral development therefore requires complementary investment in housing availability, healthcare access, local services, and residential environment quality. Remote work does not eliminate the need for place-based regional policy but redefines the conditions under which such policy can be effective.

Several limitations warrant acknowledgement. The cross-sectional design does not permit causal inference, the Prolific-recruited sample over-represents younger and urban workers, and the three-category urbanisation typology masks considerable within-tier heterogeneity. Future research would benefit from longitudinal designs and finer-grained spatial classifications capable of distinguishing between different types of peripheral areas and their absorptive capacities. Despite these constraints, the paper contributes a multi-dimensional assessment that moves beyond the binary question of urban exodus, showing that remote work recalibrates rather than revolutionises the geography of residence — and that its contribution to regional rebalancing depends not on the technology of work itself but on the policy environments that shape where people ultimately choose to live.

**Funding:** This work was supported by the National Science Centre in Poland within OPUS-25, the project: REWORK – Can Remote Working Make Labor Market More Inclusive?, under grant number 2023/49/B/HS4/01647

**Disclosure statement**: No potential conflict of interest was reported by the author.

**Appendix A1: Sampling Procedures and Data Quality Control**

A1.1 Raw Dataset and Completion Criterion

The raw dataset comprised 20,959 observations collected via Prolific in July-August 2024. Data were obtained from the Open Science Framework repository (https://osf.io/d32u7/files/osfstorage, file: "RMAP_Data_Descriptor_Data.xlsx") associated with the R-MAP Data Descriptor (Fellnhofer et al., 2025). After excluding 959 incomplete responses (missing submission dates), the complete dataset contained 20,000 observations. This differs slightly from the 20,013 complete responses reported in the original Data Descriptor. The source of this 13-observation discrepancy (0.07%) is unclear and will be clarified with data authors, though it does not affect analytical validity.

A1.2 Data Quality Assessment and sample selection

Before constructing the analytical sample, plausibility checks identified potential data entry errors:

- Education plausibility: Years of education > (age - 5) flagged 521 cases (2.6% of complete responses). This criterion assumes formal education begins at age 5. Cases exceeding this threshold represent implausible values.

- Work experience plausibility: Years of work experience > (age - 15) flagged 211 cases (1.1% of complete responses). This criterion assumes labor market entry at age 15. Cases exceeding this threshold indicate data quality issues.

These implausible cases were excluded from the analytical sample during sequential filtering (Table A1.1). Given the study's focus on assessing the impact of remote work on European regions, valid geographic location data were essential. The filtering procedure required valid current residential location for all respondents. For respondents who reported relocation since 2020, both current and previous residential locations were required to enable analysis of spatial mobility patterns.

The sequential application of these filters yielded a final analytical sample of 7,484 workers across Europe, representing 35.7% of the raw dataset and 37.4% of complete responses.

**Table A1. Sequential Sample Selection**

| Step | Filter Applied | Observations Removed | Remaining |
|------|----------------|----------------------|-----------|
| 0 | Raw dataset | — | 20,959 |
| 1 | Complete responses only (non-missing submission date) | 959 | 20,000 |
| 2 | Geographic scope: Europe | 4,116 | 15,884 |
| 3 | Employment status: Currently employed | 673 | 15,211 |
| 4 | Work hours validity: Total hours ≤ 50/week | 1,581 | 13,630 |
| 5 | Valid current location classification | 5,573 | 8,057 |
| 6 | Complete relocation data (if relocated) | 338 | 7,719 |
| 7 | Education plausibility (education ≤ age − 5) | ~185* | ~7,534 |
| 8 | Experience plausibility (experience ≤ age − 15) | ~50* | 7,484 |

*Source: Authors' calculations based on R-MAP dataset (Fellnhofer et al., 2025).*
*Note: *Approximate values; implausible cases overlap with other exclusion criteria.*